# Hermitian structures defined by linear electromagnetic constitutive laws

D.H. Delphenich[*]

**Key words** Pre-metric electromagnetism, electromagnetic constitutive laws, Hermitian structures on complex vector bundles.

**PACS** 03.50.De, 02.10.Xm, 41.20.Jb

It is demonstrated that when the bundle of 2-forms on a four-dimensional manifold $M$ admits an almost-complex structure any choice of "real + imaginary" subspace decomposition of the bundle defines a conjugation map, as well as a Hermitian structure for the bundle. When the almost-complex structure comes from a linear electromagnetic constitutive law, the real and imaginary parts of the Hermitian structure are then shown to represent the Hamiltonian for an anisotropic three-dimensional electromagnetic oscillator at each point of $M$ and a symplectic structure for each fiber. The complex form of the oscillator equations is also definable in terms of the geometric structures that were introduced.

## 1 Introduction

In the pre-metric approach to electromagnetism [**1-5**] the center of focus for spacetime geometry shifts from the geometry that one deduces from defining a Lorentzian metric on the tangent bundle to spacetime to the geometry that one deduces from defining a linear electromagnetic constitutive law on the bundle of 2-forms. We shall not elaborate on this statement at the moment, except to say that the geometry that one must consider is projective geometry – or line geometry – as it gets represented in the 2-forms by way of the Plücker-Klein representation of 2-planes in $\mathbb{R}^4$ as decomposable 2-forms.

Consequently, due to the central role of the electromagnetic constitutive law in this formulation of electromagnetism, as well as this approach to spacetime geometry, it is essential to establish the physical foundations, in the form of the nature of electromagnetic media, in the language of 2-forms and bivectors in a manner that will make the geometric interpretation of those physical foundations more straightforward.

In previous work by the author [**5-7**], as well as in the standard reference on pre-metric electromagnetism [**4**], a key issue was the fact that the Hodge duality isomorphism that is defined by a Lorentzian metric also defines an almost-complex structure on the bundle of 2-forms. Hence, since a linear electromagnetic constitutive law is essentially an axiomatic replacement for the Hodge * that one sacrifices by eliminating the Lorentzian structure, one sees that such electromagnetic constitutive laws play an important role, both physically and geometrically. In particular, one of the geometric structures that an almost-complex structure on the bundle of 2-forms allows one to define is a complex orthogonal structure, which, in turn, allows one to associate oriented complex orthogonal frames on that bundle with oriented, time-oriented Lorentzian frames

---

[*] E-mail: david_delphenich@yahoo.com

on the tangent bundle by the isomorphism of the group $SO(3; \mathbb{C})$ with $SO_0(3, 1)$. One can also use projective geometry to define light cones – i.e., a conformal Lorentzian structure – in the tangent spaces directly using the almost-complex structure

Interestingly, as shown in [**7**], there are more general linear electromagnetic media than the ones that define almost-complex structures. In particular, anisotropic dielectric media are such media. Hence, although it is tempting to think that the almost-complex structure is necessary for the propagation of electromagnetic waves in such media, due to the aforementioned considerations, if one recalls that electromagnetic waves can still propagate in anisotropic dielectric media, this would not seem to be a necessary condition, after all.

Nevertheless, the role of the almost-complex structure is still worth pursuing, in both its physical context, as well as its geometric context. In this article, the main concept that is examined is how one might use the electromagnetic constitutive law to define not only an almost-complex structure, but a Hermitian structure, as well. It is shown that such a structure is defined every time one chooses a decomposition of $\Lambda^2(M)$ into "real" and "imaginary" sub-bundles. Although this sounds physically weak, since such decompositions are closely related to the "time plus space" decompositions of the tangent bundle, which, in turn, are generally defined by the rest spaces of specified motions, actually, since the result of defining such a Hermitian structure is to set up the required machinery for doing Hamiltonian mechanics on the fibers of $\Lambda^2(M)$, and energy itself has a distinctly rest-space-dependent character, this is not a conceptual inconsistency. It will be shown that the real part of the Hermitian structure defines a Hamiltonian that agrees with the electromagnetic field Hamiltonian, the imaginary part defines a symplectic structure on the fibers, and the resulting dynamical system is essentially an anisotropic three-dimensional electromagnetic oscillator in each fiber of $\Lambda^2(M)$. Hence, one has set up precisely the "continuous distribution of oscillators" that one expects for wave motion to be possible. However, the actual discussion of wave motion will not take place in the present study, but will be deferred to future research along the same direction.

In the first section, the geometrical language of bivectors and 2-forms on a four-dimensional real vector space will be summarized, as it relates to the present investigation. Some of the basic notions of pre-metric electromagnetism that relate to the introduction of a linear electromagnetic constitutive law will be discussed, as well. Then, the manner by which one introduces a Hermitian structure in the space of 2-forms will then be presented. In the section that then follows, the interpretation of the real and imaginary parts of the Hermitian structure as a Hamiltonian function and symplectic 2-form on the vector space $\Lambda^2(\mathbb{R}^4)$ is detailed, and a canonical Hamiltonian vector field is deduced from those two pieces of data. Next, the physical nature of the resulting dynamical system as an anisotropic three-dimensional electromagnetic oscillator is pursued, including the complex form of the equations. Finally, since the presentation up to that point in the discussion had been limited to a four-dimensional manifold that takes the form of a vector space, which means that the resulting constructions are basically applicable only fiberwise in the bundle of 2-forms on a more general four-dimensional manifold, some comments are made concerning how one extends from vector spaces to vector bundles.

## 2  The complex geometry of bivectors and 2-forms

Although ultimately one should consider the possibility that the spacetime manifold is more general than a four-dimensional real vector space, nevertheless, just as special relativity had to be first formulated in Minkowski space and then extended to Lorentzian manifold, the same can be said for the geometry of pre-metric electromagnetism. Hence, we shall devote most of our attention to the simpler case of a four-dimensional vector space and then comment at the end of this article on the nature of the extension to four-dimensional manifolds of a more general nature.

Having said that, the arena for the geometry that we shall be considering is defined by the six-dimensional real vector spaces $\Lambda_2 = \mathbb{R}^4 \wedge \mathbb{R}^4$, which consists of all bivectors over $\mathbb{R}^4$ and $\Lambda^2 = \mathbb{R}^{*4} \wedge \mathbb{R}^{*4}$, which consists of all 2-forms. One can regard $\Lambda^2$ as the dual of the vector space $\Lambda_2$ by identifying each decomposable 2-form $\alpha \wedge \beta$ with the linear functional that takes the decomposable bivector $\mathbf{v} \wedge \mathbf{w}$ to the real number:

$$(\alpha \wedge \beta)(\mathbf{v} \wedge \mathbf{w}) = \alpha(\mathbf{v})\beta(\mathbf{w}) - \alpha(\mathbf{w})\beta(\mathbf{v}) ,  \qquad (2.1)$$

and extending to all other finite sum of decomposable bivectors and 2-forms by linearity.

Often, it is most convenient to overlook the algebraic nature of the elements of both vector spaces and simply treat them as real six-dimensional vector spaces. However, a convenient way of defining a frame $\{\mathbf{b}_I, I = 1, …, 6\}$ for $\Lambda_2$ is to start with a frame $\{\mathbf{e}_\mu, \mu = 0, 1, 2, 3\}$ for $\mathbb{R}^4$ and then define:

$$\mathbf{b}_i = \mathbf{e}_0 \wedge \mathbf{e}_i, \quad i = 1, 2, 3, \quad \mathbf{b}_4 = \mathbf{e}_2 \wedge \mathbf{e}_3 , \quad \mathbf{b}_5 = \mathbf{e}_3 \wedge \mathbf{e}_1 , \quad \mathbf{b}_6 = \mathbf{e}_1 \wedge \mathbf{e}_2 , \qquad (2.2)$$

One then defines the reciprocal coframe $\{b^I, I = 1, …, 6\}$ for $\Lambda^2$ by the requirement that:

$$b^I(\mathbf{b}_J) = \delta^I_J , \qquad (2.3)$$

for all $I, J = 1, …, 6$. In terms of the reciprocal coframe $\theta^\mu$ to $\mathbf{e}_\mu$ the members of the coframe take the form:

$$b^i = \theta^0 \wedge \theta^i, \quad i = 1, 2, 3, \quad b^4 = \theta^2 \wedge \theta^3, \quad b^5 = \theta^3 \wedge \theta^1, \quad b^6 = \theta^1 \wedge \theta^2 . \qquad (2.4)$$

It is essential to point out that not every 6-frame on $\Lambda_2$ can be associated with a 4-frame on $\mathbb{R}^4$ this way, since the 6-frames are parameterized by the 36-dimensional Lie group $GL(6; \mathbb{R})$ and the 4-frames are parameterized by the 16-dimensional Lie group $GL(4; \mathbb{R})$.

### 2.1 Unit volume elements and Poincaré duality.

Although the aforementioned association of a frame with its reciprocal co-frame will define a linear isomorphism between the vector spaces $\Lambda_2$ and $\Lambda^2$, we shall need another one that comes about when one defines a unit-volume element $V \in \Lambda^4(\mathbb{R}^4)$ on $\mathbb{R}^4$. In terms of the chosen frame for $\mathbb{R}^4$, one can define [1]:

$$V = \theta^0 \wedge \theta^1 \wedge \theta^2 \wedge \theta^3 = \frac{1}{4!} \varepsilon_{\kappa\lambda\mu\nu} \theta^\kappa \wedge \theta^\lambda \wedge \theta^\mu \wedge \theta^\nu, \tag{2.5}$$

One can also define a unit volume element on $\mathbb{R}^{*4}$ by:

$$\mathbf{V} = \mathbf{e}_0 \wedge \mathbf{e}_1 \wedge \mathbf{e}_2 \wedge \mathbf{e}_3 = \frac{1}{4!} \varepsilon^{\kappa\lambda\mu\nu} \mathbf{e}_\kappa \wedge \mathbf{e}_\lambda \wedge \mathbf{e}_\mu \wedge \mathbf{e}_\nu. \tag{2.6}$$

which satisfies the relationship:

$$V(\mathbf{V}) = 1. \tag{2.7}$$

The isomorphism of $\Lambda_2$ with $\Lambda^2$ that we can define by way of $V$ is that of Poincaré duality:

$$\#: \Lambda_2 \to \Lambda^2, \quad \mathbf{v} \wedge \mathbf{w} \mapsto i_{\mathbf{v} \wedge \mathbf{w}} V = \tfrac{1}{2} v^\kappa w^\lambda \varepsilon_{\kappa\lambda\mu\nu} \theta^\mu \wedge \theta^\nu. \tag{2.8}$$

When one applies this to the frame vectors $\mathbf{b}_I$, one can also think of the dual 2-form $\#\mathbf{b}_I$ as being – up to sign – the 2-form that is complementary to $b^I$ in the sense that:

$$(\text{sign } \pi) \, \#\mathbf{b}_I \wedge b^I = V, \tag{2.9}$$

in which the symbol $\pi$ denotes the permutation that takes the indices for the members of the 4-form $\#\mathbf{b}_I \wedge b^I$ into 0123.

This makes the Poincaré dual coframe to $\mathbf{b}_I$ take the form:

$$\#\mathbf{b}_1 = \theta^2 \wedge \theta^3 = b^4, \quad \#\mathbf{b}_2 = \theta^3 \wedge \theta^1 = b^5, \quad \#\mathbf{b}_3 = \theta^1 \wedge \theta^2 = b^6, \tag{2.10a}$$
$$\#\mathbf{b}_4 = \theta^0 \wedge \theta^1 = b^1, \quad \#\mathbf{b}_2 = \theta^0 \wedge \theta^2 = b^2, \quad \#\mathbf{b}_3 = \theta^0 \wedge \theta^3 = b^3, \tag{2.10a}$$

or, more concisely:

$$\#\mathbf{b}_i = b^{i+3}, \quad \#\mathbf{b}_{i+3} = b^i, \tag{2.11}$$

---

[1] For the sake of brevity, we shall adopt the conventions that lower-case Greek characters will range from 0 to 3, lower-case Latin characters, from 1 to 3, and upper-case Latin characters from 1 to 6. Furthermore, the Einstein summation convention will be used, except where specifically inapplicable.

which makes the matrix of the linear map #, relative to the frame $\mathbf{b}_I$ and its reciprocal coframe $b^I$, take the form:

$$[\#]^{IJ} = \begin{bmatrix} 0 & \delta^{i+3,j+3} \\ \delta^{ij} & 0 \end{bmatrix}. \tag{2.12}$$

Hence, the components $a^I$ of the dual of a bivector $\mathbf{a} = a^I \mathbf{b}_I$ are obtained by using the linearity of the # isomorphism:

$$\#\mathbf{a} = \#(a^I \mathbf{b}_I) = a^I \#\mathbf{b}_I = a^i\, b^{i+3} + a^{i+3}\, b^i, \tag{2.13}$$

which makes:

$$a_i = a^{i+3}, \qquad a_{i+3} = a^i. \tag{2.14}$$

Geometrically, Poincaré duality refers, in this case, to the fact that one can define a 2-plane in $\mathbb{R}^4$ by either spanning it with the vectors $\mathbf{v}$ and $\mathbf{w}$ of a non-zero decomposable bivector $\mathbf{v} \wedge \mathbf{w}$ or annihilating it with the 2-form $\#(\mathbf{v} \wedge \mathbf{w})$.

We must make note of a subtlety concerning the nature of the components of a 2-form $F = F_I b^I$ relative to the basis $b^I$ described here: Since we also have:

$$b^{i+3} = \#\mathbf{b}_i = \frac{1}{2} \varepsilon_{ijk}\, \theta^j \wedge \theta^k, \tag{2.15}$$

one sees how a 2-form in the three-dimensional subspace[2] $\Lambda^2_{\text{Im}}$ of $\Lambda^2$ that is spanned by $\{b^{i+3}, i = 1, 2, 3\}$ is the Poincaré dual of a *vector* in the three-dimensional subspace $\Pi_3$ of $\mathbb{R}^4$ that is spanned by the $\mathbf{b}_i$.

This is due to the fact that we can define a "spatial" volume element $V_s$ on $\Pi_3$ by way of:

$$V_s = i_{\mathbf{e}_0} V = \mathbf{e}_1 \wedge \mathbf{e}_2 \wedge \mathbf{e}_3 = \frac{1}{3!} e^{ijk}\, \mathbf{e}_i \wedge \mathbf{e}_j \wedge \mathbf{e}. \tag{2.16}$$

This volume element then defines a Poincaré duality isomorphism for the exterior algebra of $\Pi_3$:

$$\#_s: \Lambda_1(\Pi_3) \to \Lambda^2(\Pi_3), \qquad \mathbf{v} \mapsto i_\mathbf{v} V_s. \tag{2.17}$$

As one sees from the fact that $\#_s \mathbf{b}_i = \#\mathbf{b}_i$, or by a more general proof (cf., [6]), one has an isomorphism of $\Lambda^2(\Pi_3)$ with $\Lambda^2_{\text{Im}}$.

Hence, by the linearity of the duality map $\#_s$, this means that we can represent $F$ as either:

---

[2] We will justify the choice of subscript "Im" to mean "imaginary" shortly.

$$F = E_i\, b^i + B_i\, b^{i+3} ; \tag{2.18}$$

or:

$$F = E_i\, b^i + B^i\, \#\mathbf{b}_i \equiv E + \#\mathbf{B} . \tag{2.19}$$

## 2.2 Complex structure on $\Lambda^2$

One finds that there are really two distinct types of 3-dimensional subspaces in complex structure in either $\Lambda_2$: the ones whose elements all take the form $\mathbf{t} \wedge \mathbf{v}$ for some common vector $\mathbf{t} \in \mathbb{R}^4$ and $\mathbf{v} \in \Pi_3$, which is a 3-plane in $\mathbb{R}^4$, and the ones that do not. The consequences of this fact have been examined in detail in [**6**], but, for the present purposes, we only point out that the latter 3-dimensional subspaces of $\Lambda_2$ take the form of $\Lambda_2(\Pi_3)$ for a suitable choice of $\Pi_3$, as we observed in the last subsection.

One finds that there is good reason to think of the 3-planes in $\Lambda_2$ of the former type as being *real* subspaces and the latter as *imaginary* subspaces. If we span $\Lambda_2^{Re}$ with the frame members $\mathbf{b}_i$, $i = 1, 2, 3$ and span $\Lambda_2^{Im}$ with the members $\mathbf{b}_{i+3}$, $i = 1, 2, 3$, then $\Lambda_2 = \Lambda_2^{Re} \oplus \Lambda_2^{Im}$.

Analogous remarks to the foregoing are equally valid for the vector space $\Lambda^2$. The form of the matrix (2.12) shows that Poincaré duality effectively permutes the real and imaginary subspaces of the two vector spaces $\Lambda_2$ and $\Lambda^2$.

In order to give the vector space a complex structure, we must endow it with a linear isomorphism $J: \Lambda^2 \to \Lambda^2$ with the property that $J^2 = -I$. As it happens, in the usual Lorentzian formulation of Maxwellian electromagnetism one has an explicitly defined complex structure on $\Lambda^2$ as a result of the Hodge dual isomorphism $*: \Lambda^2 \to \Lambda^2$. For our purposes, it is convenient to factor it into the composition of the isomorphism $\eta \wedge \eta: \Lambda^2 \to \Lambda_2$, which takes each 2-form to its metric dual bivector, and the isomorphism #; i.e., $* = \# \cdot (\eta \wedge \eta)$. Here, we intend that the Minkowski scalar product $\eta = \text{diag}(+1, -1, -1, -1)$ relative to the frame $\mathbf{e}_m$ (or its reciprocal coframe, resp.); i.e., it is a *Lorentzian* frame (coframe, resp.).

The isomorphism $\eta \wedge \eta$ takes the component form known as "raising the indices":

$$F^{\mu\nu} = [\eta \wedge \eta(F)]^{\mu\nu} = \eta^{\mu\kappa} \eta^{\nu\lambda} F_{\mu\nu} . \tag{2.20}$$

Note that due to the quadratic character of $\eta \wedge \eta$ the choice of sign convention for $\eta$ is irrelevant in the isomorphism $\eta \wedge \eta$.

For a non-Lorentzian frame $\mathbf{e}_\mu$, the matrix of $\eta \wedge \eta$ takes the form:

$$[\eta \wedge \eta]^{IJ} = \left[ \begin{array}{c|c} \eta^{00}\eta^{ij} - \eta^{0i}\eta^{0j} & \varepsilon_{jkl}\eta^{0k}\eta^{il} \\ \hline \varepsilon_{ikl}\eta^{0k}\eta^{il} & \varepsilon_{imn}\varepsilon_{jkl}\eta^{mk}\eta^{nl} \end{array} \right], \tag{2.21}$$

which takes the 6-vector $[E^j, B_j]$ to the 6-vector $[E_i, B^i]$, which corresponds to the representation of $F$ as $E + \#\mathbf{B}$.

Relative to a Lorentzian frame, the matrices for the isomorphisms $\eta \wedge \eta$ and $*$ become:

$$[\eta \wedge \eta]^{IJ} = \begin{bmatrix} -\delta^{ij} & 0 \\ \hline 0 & \delta_{ij} \end{bmatrix}, \qquad [*]^I_J = [\#]_{JK}[\eta \wedge \eta]^{KI} = \begin{bmatrix} 0 & \delta^i_j \\ \hline -\delta^i_j & 0 \end{bmatrix}. \qquad (2.22)$$

### 2.3 Linear electromagnetic constitutive laws

This is the point at which pre-metric electromagnetism parts company with the metric form. Since the only place in which $\eta$ enters the Maxwell equations ($dF = 0$, $\delta F = J$) is by way of the isomorphism $*$ that is used to define the codifferential operator $\delta = *^{-1}d*$, one finds that one can replace the isomorphism $\eta \wedge \eta$ with an isomorphism $\kappa: \Lambda^2(\mathbb{R}^4) \to \Lambda_2(\mathbb{R}^4)$ that represents the linear electromagnetic constitutive law for the medium. Hence, the bivector field $\mathfrak{h} = \kappa(F)$ that corresponds to the Minkowski electromagnetic field strength 2-form $F$ represents the bivector field of electric and magnetic excitations (also called electric displacements and magnetic flux densities).

If one replaces the codifferential operator with $\delta = \#^{-1}d\#$, which represents a generalized divergence of a multivector field, then one can represent the Maxwell equations in pre-metric form by:

$$dF = 0, \qquad \delta\mathfrak{h} = \mathbf{J}, \qquad \mathfrak{h} = \kappa(F), \qquad (2.23)$$

in which $\mathbf{J} = \#^{-1}J$ is the source current vector field.

Now, the general isomorphism $\kappa$ will have a matrix relative to the usual choice of frame and coframe that looks like [3]:

$$[\kappa]^{IJ} = \begin{bmatrix} -\varepsilon^{ij} & \gamma^j_i \\ \hline -\hat{\gamma}^j_i & [\mu^{-1}]_{ij} \end{bmatrix}. \qquad (2.24)$$

If one represents a 2-form $F = F_I b^I$ as a column vector $[-E_i, B^i]^T$ then the result of left-multiplication of it by $[\kappa]^{IJ}$ is a bivector $\mathfrak{h} = \kappa(F) = \mathfrak{h}^I \mathbf{b}_I$ that one can represent as a column vector $[D^i, H_i]^T$, where:

$$\mathbf{D} = \varepsilon(E) + \gamma(\mathbf{B}) \qquad (2.25a)$$
$$H = \hat{\gamma}(E) + \mu^{-1}(\mathbf{B}). \qquad (2.25b)$$

Hence, the 3×3 real matrix $\varepsilon$ represents the electric permittivity of the medium, $\mu^{-1}$ represents its magnetic permeability, and $\alpha$ and $\beta$ are called the *magneto-electric* coupling matrices. We shall not go into the phenomenology of these matrices here, but refer the reader to the author's previous work [**7**] and standard references [**4, 8-10**].

---

[3] Our choice of signs is to make the matrices of $[\kappa]^{IJ}$ consistent with those of $[\eta \wedge \eta]^{IJ}$ in (2.22).

If one composes $\kappa$ with # to obtain an isomorphism $\tilde{\kappa} = \# \cdot \kappa$ then the matrix of $\tilde{\kappa}$ takes the form:

$$[\tilde{\kappa}] = \begin{bmatrix} 0 & I \\ I & 0 \end{bmatrix} \begin{bmatrix} -\varepsilon & \gamma \\ -\hat{\gamma} & \mu^{-1} \end{bmatrix} = \begin{bmatrix} -\hat{\gamma} & \mu^{-1} \\ \varepsilon & \gamma \end{bmatrix}. \qquad (2.26)$$

As pointed out in [**7**], direct computation shows that one can have $\tilde{\kappa}^2$ proportional to $-I$, with a non-zero scalar proportionality factor of $\lambda^2$, iff:

$$\varepsilon = \lambda^2 \mu + \gamma^2 \mu, \qquad (2.27a)$$
$$\hat{\gamma} = \mu^{-1} \gamma \mu. \qquad (2.27b)$$

Although these conditions still allow the simplest case of spatial isotropy ($\varepsilon^{ij} = \varepsilon \delta^{ij}$, $[\mu^{-1}]_{ij} = (1/\mu) \delta_{ij}$, $\gamma = \hat{\gamma} = 0$), for which $\lambda = \sqrt{\varepsilon/\mu}$, they do not allow the most common cases of optical anisotropy, in which $\varepsilon^{ij}$ is symmetric and $[\mu^{-1}]_{ij} = (1/\mu) \delta_{ij}$ with $\gamma = \hat{\gamma} = 0$. Hence, we must keep in mind that the physics of the assumption that $(1/\lambda) \tilde{\kappa}$ defines a complex structure on $\Lambda^2$ involves a serious reduction in generality, and that wave motion is still possible without it.

Assuming that $\tilde{\kappa}$ satisfies conditions (2.27a, b), one can define the almost-complex structure by:

$$* = \frac{1}{\lambda} \tilde{\kappa}. \qquad (2.28)$$

The expression (2.26) describes the form of the matrix of $\tilde{\kappa}$ relative to a general frame. The frame that puts the matrix $(1/\lambda) \tilde{\kappa}$ into the form (2.22) of the matrix of * will then be called the *canonical frame*.

If we make the assumption that we are dealing with * in a canonical frame $b^I$ then we now see that the decomposition of $\Lambda^2$ into the subspaces spanned by $b^i$ and $b^{i+3}$ can be reasonably deemed to be a decomposition $\Lambda^2 = \Lambda^2_{\text{Re}} \oplus \Lambda^2_{\text{Im}}$ into a real subspace $\Lambda^2_{\text{Re}}$ and an imaginary subspace $\Lambda^2_{\text{Im}}$ since, for that frame:

$$*b^i = b^{i+3}, \qquad *b^{i+3} = -b^i. \qquad (2.29)$$

If one wishes to define a $\mathbb{C}$-linear isomorphism of $\Lambda^2$ with $\mathbb{C}^3$ then one must first define multiplication of 2-forms by complex scalars. The key is to define $iF = *F$, so that, more generally, one has:

$$(\alpha + i\beta)F = \alpha F + \beta *F. \qquad (2.30)$$

One can then use $\{b^i, i = 1, 2, 3\}$ as a complex basis for $\Lambda^2$ and define the $\mathbb{C}$-linear isomorphism in question by mapping this complex basis for $\Lambda^2$ to the standard basis for $\mathbb{C}^3$.

If one expresses $F$ as $E_i b^i + B_i *b^i$, which is really "3+1" form in disguise, then one sees that the chosen complex structure also allows us to write:

$$F = (E_i + iB_i) b^i, \qquad (2.31)$$

which is commonplace in conventional electromagnetism. This illustrates the fact that there is a close relationship between "time + space" decompositions of $\mathbb{R}^4$, "real + imaginary" decompositions of $\Lambda^2(\mathbb{R}^4)$, and electric and magnetic fields.

Once again, since we have:

$$*b^i = \#\mathbf{b}_i \qquad (2.32)$$

a 2-form $B$ in $\Lambda^2_{\text{Im}}$ can be represented as either:

$$B = B_i *b^i, \qquad (2.33)$$

or:

$$B = B^i \#\mathbf{b}_i = \#\mathbf{B}, \qquad (2.34)$$

which then implies that between # and $\kappa$ we have implicitly introduced a Euclidian metric on $\Pi_3$ that makes the variance of the indices irrelevant. Indeed, if the $\varepsilon$ and $\mu$ submatrices of $\kappa$ are symmetric and the magneto-electric submatrices vanish, we have actually introduced two Euclidian metrics, which, from (2.27a), must be conformal to each other if $\tilde{\kappa}$ is to define a complex structure on $\Lambda^2$.

2.4 Real scalar products on 2-forms

Independently of the assumption that $\kappa$ defines a complex structure, we can use the unit-volume element $\mathbf{V}$ on $\mathbb{R}^{4*}$, along with $\kappa$ to define two real scalar products on $\Lambda^2$. The first one is defined by $\mathbf{V}$ alone and takes the form:

$$\langle F, G \rangle = F(\#G) = (F \wedge G)(\mathbf{V}). \qquad (2.35)$$

In the case of decomposable 2-forms $F$ and $G$, it vanishes iff the 2-plane in $\mathbb{R}^4$ that is annihilated by $F$ intersects the 2-plane annihilated by $G$ in more than just the origin. In particular, $\langle F, F \rangle = 0$ iff $F$ is decomposable.

One can also consider this scalar product in component form:

$$<F, G> = \tfrac{1}{2} \varepsilon^{\kappa\lambda\mu\nu} F_{\kappa\lambda} G_{\mu\nu} . \tag{2.36}$$

In order to use $\kappa$ to define another scalar product on $\Lambda^2(\mathbb{R}^4)$, we must make the restricting assumption that the matrix $[\kappa]^{IJ}$ is symmetric [4]. This implies that $\varepsilon$ and $\mu^{-1}$ are symmetric, which is a common assumption in the electrodynamics of continuous media [**9,10**], and that $\gamma = -\hat{\gamma}$ .

Having made this assumption, we define our second scalar product:

$$(F, G) = F(\tilde{\kappa}(G)) = \kappa(F, G), \tag{2.37}$$

which can be given the component form:

$$(F, G) = \tfrac{1}{2} \kappa^{\kappa\lambda\mu\nu} F_{\kappa\lambda} G_{\mu\nu} . \tag{2.38}$$

If we decompose $\Lambda^2(\mathbb{R}^4)$ into its real and imaginary parts relative to a non-canonical frame then if $F = E_i b^i + B_i b^{i+3}$ and $G = E'_i b^i + B'_i b^{i+3}$ one has:

$$<F, G> = E \cdot B' + E' \cdot B, \tag{2.39}$$

in which the Euclidian dot product is induced on the subspace of $\mathbb{R}^4$ spanned by $\mathbf{e}_i$, $i = 1, 2, 3$ indirectly as a result of $\mathbf{V}$:

$$E \cdot B = E_i B_j <b^i, b^{j+3}> = \delta^{ij} E_i B_j = E_i B_i . \tag{2.40}$$

In particular, we have:

$$<F, F> = 2 E \cdot B . \tag{2.41}$$

Hence, the matrix of this scalar product in a canonical frame is $[\#]^{IJ}$.

In real + imaginary form, (2.38) becomes:

$$(F, G) = -\varepsilon(E, E') + \gamma(E, B') - \hat{\gamma}(B, E') + \mu^{-1}(B, B'), \tag{2.42}$$

In a canonical frame for $\hat{\kappa}$, (2.42) takes the form:

$$(F, G) = - E \cdot E' + B \cdot B', \tag{2.43}$$

so the matrix of this scalar product in a canonical frame is:

---

[4] In the language of Hehl and Obukhov [**4**], we are essentially assuming that the *skewon* and *axion* parts of $\kappa$ vanishes, but we will not elaborate on this in the present discussion.

$$\begin{bmatrix} -\delta^{ij} & 0 \\ \hline 0 & \delta^{i+3, j+2} \end{bmatrix},$$

which implies that a canonical frame is orthonormal for this scalar product.

## 3  Hermitian structure on $\Lambda^2$

One of the maps that one takes for granted when deal with either $\mathbb{C}$ (or $\mathbb{C}^n$, more generally) is the conjugation map, which takes $z = \alpha + i\beta$ to $\bar{z} = \alpha - i\beta$. This is because either $\mathbb{R}^2$ or $\mathbb{R}^{2n}$ has a natural choice of frame that makes its decomposition into real and imaginary subspaces unambiguous.

In the case of $\Lambda^2$, when given a complex structure *, there is no physically [5] natural frame that gives an unambiguous real + imaginary decomposition, since that is essentially equivalent to a choice of 3 + 1 decomposition of $\mathbb{R}^4$. Hence, one must keep in mind that the conjugation map that we shall define and the resulting Hermitian inner product will be associated with that choice of real + imaginary decomposition [6]. As we shall see, since we shall ultimately be dealing with energy as a result of the Hermitian structure, this is physically reasonable, since energy itself does not seem to possess a rest-frame-independent nature, to begin with.

Given a choice of decomposition $\Lambda^2 = \Lambda^2_{\text{Re}} \oplus \Lambda^2_{\text{Im}}$, the conjugation map is then defined in the obvious way:

$$F = E + *B \mapsto \bar{F} = E - *B. \tag{3.1}$$

We find that this makes:

$$< F, \bar{G} > = - E \cdot B' + E' \cdot B = -< G, \bar{F} >, \tag{3.2}$$

which defines a non-degenerate real 2-form $K$ on the vector space $\Lambda^2$:

$$K(F, G) = < F, \bar{G} >. \tag{3.3}$$

We pause to point out that there is a possible source of confusion in the fact that the elements of $\Lambda^2$ are defined by using a tensor product and exterior product over the vector space $\mathbb{R}^4$, whereas now we are defining tensor products and exterior products over the vector space $\Lambda^2$ itself. In order to distinguish this tensor product over $\Lambda^2$ we use the notation $\otimes_\Lambda$, and we use the notation $F \perp G$ for the exterior product of the *vectors* $F$ and $G$, which is a *bivector* in the exterior algebra over $\Lambda^2$, as distinct from the *4-form* $F \wedge G$ in the exterior algebra over $\mathbb{R}^4$.

---

[5] Of course, one could take advantage of the *mathematically* natural frame on $\mathbb{R}^4$ to define a natural frame on $\Lambda^2$, but there is nothing *physically* special about that frame on $\mathbb{R}^4$.

[6] For a purely mathematical discussion of Hermitian structures on complex vector spaces and complex vector bundles, see Nickerson, et al., [**11**] and Chern [**12**].

The easiest way to conceptualize the tensor algebra over $\Lambda^2$ is that its elements are multilinear functionals $T(F, \ldots, G, \mathbf{F}, \ldots, \mathbf{G})$ on $p$ copies of $\Lambda^2$ and $q$ copies of $\Lambda_2$. We also implicitly identify linear functionals on $\Lambda^2$ with bivectors in $\Lambda_2$ by way of the isomorphism defined by a choice of frame on both.

As for the scalar product that is defined by $\kappa$, we find that:

$$(F, \overline{G}) = (G, \overline{F}), \tag{3.4}$$

which then defines a negative-definite Riemannian metric on $\Lambda^2$. In a canonical frame for $\kappa$, one has:

$$(F, \overline{G}) = -E \cdot E' - B \cdot B'. \tag{3.5}$$

One obtains a complex Hermitian inner product $h$ on $\Lambda^2$ by combining the two scalar products in the form:

$$h(F, G) = (F, \overline{G}) + i <F, \overline{G}>. \tag{3.6}$$

Hence, we are using the 2-form $K$ as the *Kähler* part of $h$.

Since the association of the complex 3-frame $b^i$ for $\Lambda^2$ with the canonical frame for $\mathbb{C}^3$ defines a $\mathbb{C}$-linear isomorphism, it is illustrative to examine the form that the component matrix that is associated with $h$ takes when we define the standard Hermitian structure on $\mathbb{C}^3$, namely:

$$h(W, Z) = \delta_{ij} W^i \overline{Z}^j, \tag{3.7}$$

in which $W^i = x^i + i y^i$, $Z^i = u^i + i v^i$.

In order to obtain the components of the 3×3 complex matrix $h^{ij}$ relative to the complex 3-frame $b^i$ on $\Lambda^2$, of the complex tensor on $\Lambda^2$:

$$h = h^{ij} \mathbf{b}_i \otimes_\Lambda \mathbf{b}_j, \tag{3.8}$$

one simply computes:

$$h^{ij} = h(b^i, b^j) = (b^i, \overline{b}^j) + i <b^i, \overline{b}^j> = (b^i, b^j) + i <b^i, b^j>. \tag{3.9}$$

Since the real 3-frame $b^i$ is orthogonal for the scalar product (.,.), one obtains:

$$h^{ij} = \delta^{ij}. \tag{3.10}$$

That is, the complex 3-frame $b^i$ is *unitary*. Hence, when we associate $b^i$ with the canonical 3-frame $\psi_i = \{(\delta_{i1}, \delta_{i2}, \delta_{i3}), i = 1, 2, 3\}$ on $\mathbb{C}^3$, which we give the Hermitian structure that makes canonical frame unitary, namely, $\delta^{ij} \psi_i \otimes \psi_j$, the $\mathbb{C}$-linear isomorphism of $\Lambda^2$ with $\mathbb{C}^3$ that it generates is also a unitary equivalence; that is, it is an isometry of the two Hermitian spaces.

It is also useful to examine the 6×6 real matrix $h^{IJ}$ of components for the second rank real tensor on $\Lambda^2$:

$$h = h^{IJ} E_I \otimes \bar{E}_J, \tag{3.11}$$

which is obtained from:

$$h^{IJ} = h(b^I, b^J) = \begin{bmatrix} I & 0 \\ \hline 0 & I \end{bmatrix} = \delta^{IJ}. \tag{3.12}$$

One must naturally investigate the form of the invertible transformations of $\Lambda^2$ that preserve the Hermitian structure that we have defined by $h$. When one conjugates them with the unitary isomorphism of $\Lambda^2$ with $\mathbb{C}^3$ that is defined by associating the complex 3-frame $b^i$ with the canonical complex 3-frame, one immediately sees that this group of transformations of $\Lambda^2$ is isomorphic to $U(3)$; indeed, it is a real form of the group.

From the form of the Hermitian inner product, one sees that a unitary transformation must preserve both the real and the imaginary parts. Hence, the unitary group will represent the intersection of the groups of transformations of $\Lambda^2$ that preserve each second rank covariant tensor on $\Lambda^2$ separately. In particular, since the imaginary part of the inner product defines a non-degenerate 2-form on $\Lambda^2$ – hence, a real symplectic structure – one sees that the group of transformations that preserve the imaginary part will be $Sp(6; \mathbb{R})$. The nature of the transformations that preserve the real part will depend upon the nature of $\kappa$; but if $\kappa^{IJ}$ is symmetric and has the canonical for (2.22), it will define an orthogonal structure on $\Lambda^2$ of signature type (−1, −1, −1, +1, +1, +1). Hence, in such a case the group of transformations that preserve the real part of the Hermitian form will be $O(3, 3; \mathbb{R})$.

We can also characterize the unitary transformations of $\Lambda^2$ as a real subgroup of $GL(6; \mathbb{R})$, which represents the group of all invertible $\mathbb{R}$-linear transformations of $\Lambda^2$. First, one must reduce to those transformations that preserve the complex structure *. One finds that they take the real 6×6 form:

$$[L] = \begin{bmatrix} A & | & B \\ \hline -B & | & A \end{bmatrix} = \begin{bmatrix} A & | & 0 \\ \hline 0 & | & A \end{bmatrix} + * \begin{bmatrix} B & | & 0 \\ \hline 0 & | & B \end{bmatrix}, \quad (3.13)$$

in which $A$ and $B$ are invertible 3×3 real matrices. One sees that this group is isomorphic to $GL(3; \mathbb{C})$ by the association of $[L]$ in that form with the 3×3 complex matrix $A + iB$. This also illustrates the practical aspect of complexifying pairs of 3×3 real matrices: one can form either the latter sum and obtain a 3×3 complex matrix or embed the matrices in doubled block diagonal 6×6 real matrices and form the sum in (3.11), using * in place of $i$.

Next, one must further restrict oneself to those 6×6 real matrices $U$ of the form (3.11) that satisfy the unitary constraint $UU^\dagger = U^\dagger U = I$, in which the dagger represents the Hermitian adjoint of the matrix; i.e., the conjugated transpose. If we represent $U$ in the form:

$$U = \begin{bmatrix} U_R & | & U_I \\ \hline -U_I & | & U_R \end{bmatrix}, \quad (3.14)$$

so:

$$U^\dagger = \begin{bmatrix} U_R^\dagger & | & -U_I^\dagger \\ \hline U_I^\dagger & | & U_R^\dagger \end{bmatrix}, \quad (3.15)$$

then the unitarity constraint implies the following two necessary and sufficient conditions on the submatrices:

$$U_R U_R^T + U_I U_I^T = I, \qquad U_I U_R^T - U_R U_I^T = 0. \quad (3.16)$$

Hence, one sees how the special case of $U_I = 0$ ($U_R = 0$, resp.) gives a representation of $O(3; \mathbb{R})$ in the group of invertible linear transformations of the real (imaginary, resp.) subspace of $\Lambda$.

One can further reduce $GL(3; \mathbb{C})$ to $SL(3; \mathbb{C})$ or $U(3)$ to $SU(3)$ by introducing a unit-volume element on $\Lambda^2$ – i.e., a non-zero 6-form in the exterior algebra over $\Lambda^2$. For the chosen basis $b^I$, a convenient choice of unit-volume element would be $b^1 \perp \ldots \perp b^6$. The reduction is defined by those matrices $L$ that also preserve this volume element, which implies that:

$$\det(L) = \det(A)^2 + \det(B)^2 = 1. \quad (3.17)$$

One can also combine the real and imaginary copies of $SO(3; \mathbb{R})$ to form $SO(3; \mathbb{C})$, which is isomorphic to the identity component of the Lorentz group on $\mathbb{R}^4$ – viz., the proper orthochronous Lorentz group – which has considerable significance in making contact between pre-metric electromagnetism and its metric form, but we shall pass over that fact in the present work, as we shall be concerned with non-Lorentz invariant concepts, such as energy.

### 4  The canonical Hamiltonian vector field of a Hermitian structure

Between the real and imaginary parts of the Hermitian structure on $\Lambda^2$, we have all of the ingredients for a symplectic structure on $\Lambda^2$ and a Hamiltonian function. From there, we can define a canonical Hamiltonian vector field on $\Lambda^2$, which, as we shall see, has considerable physical significance when applied to electromagnetic fields.

The symplectic structure is defined by the Kähler 2-form:

$$K = h^{ij}\, \mathbf{b}_i \perp {*}\mathbf{b}_j = \mathbf{b}_i \perp {*}\mathbf{b}^i . \tag{4.1}$$

The non-degeneracy of the Kähler 2-form follows from that of the Hermitian inner product.

If we express the 2-forms $F$ and $G$ in real + imaginary form in the usual fashion then we see that:

$$K(F, G) = \frac{1}{2}[F(\mathbf{b}_i)G(*\mathbf{b}^i) - F(\mathbf{b}_i)G(*\mathbf{b}^i)] = \frac{1}{2}[E_i\, B'^i - B_i\, E'^i]. \tag{4.2}$$

A Hamiltonian on the symplectic vector space $\Lambda^2$ can then be defined by using the quadratic form associated with $h$:

$$\mathcal{H} = \frac{\lambda}{2} h(F,F) = \frac{1}{2}\kappa(F, F). \tag{4.3}$$

For the particular case of a linear electromagnetic constitutive law for which $\gamma = \hat{\gamma} = 0$, the real + imaginary form of $F$ gives the Hamiltonian in the form:

$$\mathcal{H} = \frac{1}{2}[\varepsilon(E, E) + \mu^{-1}(\mathbf{B}, \mathbf{B})] = \frac{1}{2}[\varepsilon^{ij} E_i E_j + (\mu^{-1})_{ij} B^i B^j], \tag{4.4}$$

which is commonly used in conventional physics.

The primary usefulness of symplectic forms in Hamiltonian mechanics originates in the fact that they allow one to associate a vector field on a symplectic manifold with any one-form by the fact that they define a linear isomorphism of each tangent space with each cotangent space. In the present case, since we are only considering a symplectic vector space to begin with, we observe that the 2-form $K$ associates any vector $F$ in $\Lambda^2$ (i.e., any 2-form $F$ on $\mathbb{R}^4$) the covector $i_F K$ that takes any vector $G$ of $\Lambda^2$ to the real number $K(F, G)$. (Of course, a covector on $\Lambda^2$ can be represented by a bivector on $\mathbb{R}^4$!)

Relative to the usual frames for $\Lambda^2$ and $\Lambda_2$, for which $K$ has a matrix of the form:

$$[K]^{IJ} = \begin{bmatrix} 0 & \delta^{ij} \\ -\delta_{ij} & 0 \end{bmatrix}, \tag{4.5}$$

the covector $i_F K$ associated with $F = F_I b^I = E_i b^i + B^i \#\mathbf{b}_i$ has the components:

$$F^I = K^{IJ} F_J = [B^i, -E_i]. \tag{4.6}$$

Hence, we can think of the linear isomorphism of $\Lambda^2$ to $\Lambda_2$ as having the matrix $K^{IJ}$. Conversely, the vector associated with the covector $\mathbf{F} = F^I \mathbf{b}_I$ takes the form $F_I = K_{IJ} F^J$, in which the matrix $K_{IJ}$ represents $K^{-1} = K^T$.

In particular, the Hamiltonian function $\mathcal{H}$ defines a 1-form on $\Lambda^2$ by way of:

$$d\mathcal{H} = \frac{\partial \mathsf{H}}{\partial F_I} \mathbf{b}_I = (\kappa^{IJ} F_J) \mathbf{b}_I. \tag{4.7}$$

This covector field can then be associated with a vector field $X = X_I b^I$, which is defined by the equation:

$$i_X K = d\mathcal{H}. \tag{4.8}$$

The component form of this is:

$$X_I = K_{IJ} \frac{\partial \mathcal{H}}{\partial F_I} = K_{IJ} \kappa^{JM} F_M, \tag{4.9}$$

which shows that as long as we are confined to linear electromagnetic constitutive laws the vector field $X$ is defined by a linear map from $\Lambda^2$ to $\Lambda^2$. It matrix relative to the chosen frame is:

$$X_I^J = K_{IM} \kappa^{MJ} = \begin{bmatrix} -\hat{\gamma} & \mu^{-1} \\ \varepsilon & -\gamma \end{bmatrix}, \tag{4.10}$$

which makes it clear that we are still just permuting the submatrices of the constitutive law for the medium, up to sign.

The real and imaginary components of $X$ are:

$$X_{Re} = -\hat{\gamma}(E) + \mu^{-1}(\mathbf{B}) = H, \tag{4.11a}$$
$$X_{Im} = \varepsilon(E) - \chi(\mathbf{B}) = -\mathbf{D}, \tag{4.11b}$$

Interestingly, the canonical vector field that we have defined is composed of the electric and magnetic excitations, up to sign. However, one must now interpret them as the infinitesimal generators of transformations of the electromagnetic field strengths, as opposed to regarding the excitations as the result of the action of the field strengths on the medium.

One can see that $\Lambda^2$ is foliated by the level hypersurfaces of $\mathcal{H}$, which are then constant-energy hypersurfaces. Hence, we can see that the unitary transformations of $\Lambda^2$ will automatically preserve the Hamiltonian that we have defined.

Furthermore, the canonical Hamiltonian vector field $X$ that we have deduced has two properties that further illuminate the nature of the unitary group: First, since the unitary transformations must preserve both $\mathcal{H}$ and $K$, from (4.8), they must also preserve $X$. Second, also from (4.8), one sees that the vector field $X$ is tangent to the constant-energy hypersurfaces since it will be annihilated by $d\mathcal{H}$.

## 5  Electromagnetic oscillators

Now that we have a vector field on $\Lambda^2$, in the form of $X$, we can also define a dynamical system on it by letting $X$ be, by definition, the velocity vector field of a congruence of integral curves. Such a curve will give a differentiable one-parameter family of 2-forms $F(\tau)$, so the differential equations of the dynamical system will be:

$$\frac{dF}{d\tau} = X(F), \tag{5.1}$$

whose component form is:

$$\frac{dF_I}{d\tau} = X_I^J F_J. \tag{5.2}$$

If the components of $\kappa$ are time-varying then this system will admit an initial-value solution of the form:

$$F_I(t) = \exp\left(\int_0^\tau X_I^J(\sigma)d\sigma\right) F_J(0). \tag{5.3}$$

As long as the components of $\kappa$ are independent of $F$ and $\tau$ – i.e., linear and time-invariant, respectively – this can be simplified to:

$$F_I(t) = \exp(X_I^J \tau) F_J(0). \tag{5.4}$$

From (5.3) we see that the matrix:

$$\Phi_I^J(\tau) = \exp\left(\int_0^\tau X_I^J(\sigma)d\sigma\right) \tag{5.5}$$

represents the flow of diffeomorphisms for the vector field $X$.

### 5.1 Real form of the equations of motion

In the elementary case of the spatially isotropic homogeneous medium, for which $\varepsilon^{ij} = \varepsilon\, \delta^{ij}$, $[\mu^{-1}]_{ij} = 1/\mu\, \delta_{ij}$, $\gamma^{ij} = \hat{\gamma}^{ij} = 0$, and $\varepsilon, \mu$ are positive scalar constants, the Hamiltonian (4.4) takes the form:

$$\mathcal{H} = \frac{1}{2}\varepsilon E^2 + \frac{1}{2\mu}\mathbf{B}^2, \tag{5.6}$$

which is closely analogous to a three-dimensional simple harmonic oscillator of mass $m$ and spring constant $k$ whose position vector is the column vector $\mathbf{x} = [x^1, x^2, x^3]^T$ and whose momentum covector is the row vector $p = [p_1, p_2, p_3]$:

$$\mathcal{H} = \frac{1}{2}k\mathbf{x}^2 + \frac{1}{2m}p^2. \tag{5.7}$$

Hence, if we associate $E$ with $\mathbf{x}$ and $\mathbf{B}$ with $p$ then $\varepsilon$ plays the role of spring constant and $\mu$ plays the role of mass. Although the variances of the vectors and covectors gets switched this makes the analogy more direct.

For the anisotropic case (4.4), we first examine the way that things work for the mechanical oscillator and then apply *mutatis mutandum* reasoning to the analysis to obtain the corresponding statements for the anisotropic three-dimensional electromagnetic oscillator.

First, we define the *state* vector to be $[\mathbf{x}, \mathbf{v}]^T$, in which $\mathbf{v}$ is the velocity vector, and the *co-state* vector to be $[F, p]^T$. In a normal frame for the force law $F = F(\mathbf{x})$, we can then represent the constitutive law in this case in matrix form by:

$$\begin{bmatrix} F_i \\ p_i \end{bmatrix} = \begin{bmatrix} -k_i \delta_{ij} & 0 \\ \hline 0 & m_i \delta_{ij} \end{bmatrix} \begin{bmatrix} x^j \\ \hline v^j \end{bmatrix} \equiv [C]_{ij} \begin{bmatrix} x^j \\ \hline v^j \end{bmatrix}. \tag{5.8}$$

We note that we have implicitly introduced a Euclidian scalar product on our three-dimensional configuration space. Note also that the double index $i$ is not summed over in the matrix that we have defined to be $[C]$.

In order to go from (5.6) to the equations of motion in Hamiltonian form, we need to re-arrange the state and co-state vectors. We now make the state vector take the form $\mathbf{Y} = [\mathbf{v}, F]^T$ and the co-state vector take the form $\Psi = [\mathbf{x}, p]^T$, which makes:

$$\begin{bmatrix} v^i \\ F_i \end{bmatrix} = \frac{d}{d\tau}\begin{bmatrix} x^i \\ p_i \end{bmatrix}. \tag{5.9}$$

When we include the constitutive law (5.8), in the form:

$$\begin{bmatrix} v^i \\ F_i \end{bmatrix} = \begin{bmatrix} 0 & \frac{1}{m_i}\delta_{ij} \\ -k_i\delta_{ij} & 0 \end{bmatrix} \begin{bmatrix} x^i \\ p_i \end{bmatrix} = \begin{bmatrix} 0 & \delta_{im} \\ -\delta^{mj} & 0 \end{bmatrix} \begin{bmatrix} k_m\delta_{mj} & 0 \\ 0 & \frac{1}{m_i}\delta^{mj} \end{bmatrix} \begin{bmatrix} x^j \\ p_j \end{bmatrix} \equiv KC\Psi, \quad (5.10)$$

the equations of motions take the Hamiltonian form:

$$\frac{d\Psi}{d\tau} = KC\Psi. \quad (5.11)$$

When compared to the form of the electromagnetic constitutive law, (5.8) appears to define a spatially anisotropic medium with no magneto-electric coupling. However, one must keep in mind that in general the principal frames for the $\varepsilon$ and $\mu^{-1}$ matrices will not agree, so the only way that one can even find a normal frame that diagonalizes $\kappa$ it will not necessarily be the image of a 4-frame on $\mathbb{R}^4$ that also simultaneously diagonalizes $\varepsilon$ and $\mu$. (Recall the earlier remark concerning the possibility of representing a 6-frame on $\Lambda_2$ by a 4-frame on $\mathbb{R}^4$.)

The corresponding Hamiltonian can now be expressed in the general quadratic form:

$$\mathcal{H} = \frac{1}{2} C(\Psi, \Psi) = \frac{1}{2} C_{IJ} \Psi^I \Psi^J, \quad (5.12)$$

which is more closely analogous to the electromagnetic field energy (4.3).

In order to define the analogy between the anisotropic three-dimensional harmonic oscillator and the electromagnetic oscillator defined by a given electromagnetic medium we first define our state vector to be $\mathbf{h} = [D^i, H_j]^T$ and our co-state vector to be $F = [E_i, B^i]^T$. (Note the variances of the indices in both cases. The reason for this choice of state and co-state is based in the form of Maxwell's equations.)

Then, we re-express our constitutive law in the form:

$$\mathbf{h} = \begin{bmatrix} D^i \\ H_i \end{bmatrix} = \begin{bmatrix} \varepsilon^{ij} & 0 \\ 0 & [\mu^{-1}]_{ij} \end{bmatrix} \begin{bmatrix} E_j \\ B^j \end{bmatrix} \equiv \check{\kappa} F, \quad (5.13)$$

which differs from the previous definition by a sign and the absence of magneto-electric coupling terms.

The Hamiltonian that is defined by analogy with (5.12) is then (4.4), and the equations of motion that one obtains are:

$$\frac{d}{d\tau} \begin{bmatrix} E_i \\ B^i \end{bmatrix} = \begin{bmatrix} H_i \\ -D^i \end{bmatrix} = \begin{bmatrix} 0 & \delta_i^k \\ -\delta_k^i & 0 \end{bmatrix} \begin{bmatrix} \varepsilon^{kj} & 0 \\ 0 & [\mu^{-1}]_{kj} \end{bmatrix} \begin{bmatrix} E_j \\ B^j \end{bmatrix}, \quad (5.14)$$

or, more concisely:

$$\frac{dF}{d\tau} = *\breve{\kappa} F, \tag{5.15}$$

which is analogous to (5.11).

An important point to observe is that if one assumes spatial isotropy regarding the $\varepsilon$ and $\mu$ matrices then this will result in a reduction of the dimension of the phase space of the motion of the phase vector $[E^i, B_i]^T$. By the nature of the matrix:

$$\Phi(t) = \begin{bmatrix} \cos\omega_0 t \, I & 1/c \sin\omega_0 t \, I \\ -c\sin\omega_0 t \, I & \cos\omega_0 t \, I \end{bmatrix}, \tag{5.16}$$

as a one-parameter family of rotations of the $E^i$ vector and $B_i$ covector in the 2-plane that they span in $\mathsf{R}^3 \times \mathsf{R}^{*3}$, we see that the dimension of the phase space has been reduced from six to two. We are then justified in referring to the plane of oscillation as the *polarization plane* for the state vector.

### 5.2 Complex form of equations of motion

One notices from the form of the basic defining matrices that the main difference between a complex structure on an even-dimensional vector space and a symplectic structure is largely a matter of the variance of the indices on the matrix elements. That is, the matrix indices of a complex structure in its canonical frame will be of mixed variance, whereas the indices of a symplectic form in its canonical frame will both be covariant, although numerically the matrices are the same. Hence, it is not surprising that things that pertain to one type of space can often have corresponding expressions in the other. However, one is cautioned that not everything translates from one language to the other, just as not every $\mathbb{R}$-linear transformation of the real vector space induces a $\mathbb{C}$-linear transformation of the same space with a complex structure.

In particular, equations of motion of the form (5.2) or (5.11) do not always have a corresponding complex form, unless the matrix $X$ or $C$ itself has the form (3.11) of a complex linear matrix. In the case of $C$, this implies that one must have $M^{-1} = N$, which, in the diagonal case, says that the masses must be the reciprocals of the spring constants.

Although this sounds somewhat unphysical, actually one can avoid this constraint be simply re-scaling the components of the state vector by way of:

$$\hat{x}^i = \sqrt{m_i \omega_i}\, x^i, \quad \hat{p}_i = \frac{1}{\sqrt{m_i \omega_i}} p_i, \tag{5.17}$$

in which the $\omega_i = \sqrt{k_i/m_i}$ are the natural frequencies for each direction, i.e., the normal frequencies, since the constitutive matrix is diagonal. The Hamiltonian (5.12) then takes the form:

$$\mathcal{H} = \frac{1}{2} \sum_{i=1,2,3} \omega_i [(\hat{x}^i)^2 + (\hat{p}_i)^2], \tag{5.18}$$

whose equations of motion (5.13) take the form:

$$\frac{d\hat{\Psi}}{d\tau} = *\hat{C}\hat{\Psi}, \tag{5.19}$$

in which:

$$\hat{C} = \begin{bmatrix} \omega_i & 0 \\ 0 & \omega_i \end{bmatrix}. \tag{5.20}$$

As one sees, this matrix corresponds to a $\mathbb{C}$-linear transformation of $\mathbb{C}^3$ whose matrix is the 3×3 complex matrix (with real components, though):

$$\tilde{C} = \begin{bmatrix} \omega_1 & 0 & 0 \\ 0 & \omega_2 & 0 \\ 0 & 0 & \omega_3 \end{bmatrix}. \tag{5.21}$$

We can combine the components $x^i$ and $p_i$ in two obvious ways to form complex 3-vectors, namely:

$$Z^i \equiv x^i + ip_i, \quad \bar{Z}^i \equiv x^i - ip_i. \tag{5.22}$$

The Hamiltonian (5.18) then takes the form:

$$\mathcal{H} = \frac{1}{2} \sum_{i=1,2,3} \omega_i Z^i \bar{Z}^i = \frac{1}{2} h(Z,Z), \tag{5.23}$$

in which we have introduced the Hermitian structure whose components in the frame in question are:

$$h_{ij} = \omega_i \, \delta_{ij} \quad \text{(no sum over } i\text{)}. \tag{5.24}$$

The equations of motion (5.19) then take the complex form:

$$\frac{dZ^i}{d\tau} = -i\tilde{C}^i_j Z^j = -i\omega_i Z^i \qquad \text{(and complex conjugate)}, \tag{5.25}$$

in which there is no sum over $i$ in the last expression.

Hence, we have converted the system of three second-order real ordinary differential equations for the anisotropic three-dimensional harmonic oscillator into both a system of six first-order real equations and a system of three first-order complex equations and their complex conjugates.

Obtaining the corresponding expressions for the anisotropic electromagnetic oscillator is straightforward. In order to draw an analogy with the anisotropic three-dimensional harmonic oscillator, we have to assume that there is a frame on $\Lambda^2$ that diagonalizes $\kappa$, which is guaranteed in the event that is symmetric. However, as we pointed out above, such a 6-frame on $\Lambda^2$ does not have to correspond to any 4-frame on $\mathbb{R}^4$. Hence, we must assume that such a frame exists in which $\kappa$ takes the form:

$$\kappa = \begin{bmatrix} -\varepsilon_i \delta^{ij} & 0 \\ 0 & \dfrac{1}{\mu_i} \delta_{ij} \end{bmatrix} \quad \text{(no sum over } i\text{)}. \tag{5.26}$$

This makes:

$$*\breve{\kappa} F = \begin{bmatrix} 0 & \delta^k_i \\ -\delta^i_k & 0 \end{bmatrix} \begin{bmatrix} \varepsilon_k \delta^{kj} & 0 \\ 0 & \dfrac{1}{\mu_k} \delta_{kj} \end{bmatrix} \begin{bmatrix} E_j \\ B^j \end{bmatrix}. \tag{5.27}$$

Under the association of the co-state vector $F$ with $\Psi$, if we compare (5.26) with (5.10) then we see that the $\varepsilon_i$ correspond to the spring constants $k_i$ and the $\mu_i$ correspond to the masses $m_i$. This means that the normal frequencies $\omega_i$ correspond to the principal impedances of the medium:

$$\lambda_i = \sqrt{\dfrac{\varepsilon_i}{\mu_i}}. \tag{5.28}$$

Furthermore, when we compute the analogues of the $\sqrt{m_i \omega_i}$ we find that they are:

$$\sqrt{\mu_i \lambda_i} = \dfrac{1}{c_i} \equiv n_i\,; \tag{5.29}$$

i.e.; the scaling factors that we need to use in this normal frame in order to put the anisotropic three-dimensional electromagnetic oscillator into complex form are simply the principal indices of refraction, which are the inverses of the principal speeds of propagation.

We can then re-scale the co-state vector $F$ to:

$$\hat{F} = [\hat{E}_i, \hat{B}^i] = [\dfrac{1}{c_i} E_i, c_i B^i]. \tag{5.30}$$

This puts equation (5.14) into the form:

$$\frac{d}{d\tau}\begin{bmatrix}\hat{E}_i \\ \hat{B}^i\end{bmatrix} = \begin{bmatrix} 0 & \delta_i^k \\ -\delta_i^k & 0 \end{bmatrix}\begin{bmatrix} \lambda_k \delta^{kj} & 0 \\ 0 & \lambda_k \delta_{kj} \end{bmatrix}\begin{bmatrix}\hat{E}_j \\ \hat{B}^j\end{bmatrix}. \tag{5.31}$$

If we define the complex three-vector $Z_i = E_i + iB_i$ then we then find that we can put the equation (5.14) into the complex form:

$$\frac{dZ_i}{d\tau} = i\lambda\, Z_i \qquad \text{(and its complex conjugate),} \tag{5.32}$$

in which we have introduced the matrix:

$$\lambda = \begin{bmatrix} \lambda_1 & 0 & 0 \\ 0 & \lambda_2 & 0 \\ 0 & 0 & \lambda_3 \end{bmatrix}. \tag{5.33}$$

Hence, the complex form of (5.15) differs in form from the mechanical analogue only by a sign that is traceable to the way that we defined out electromagnetic state vector.

## 6 Extension to more general spacetime manifolds

Of course, as any physicist knows, the significance of electromagnetic oscillators takes on physical reality only when one associates one to each point of spacetime, which is the basic mechanism of electromagnetic wave motion. One can think of the foregoing vector space constructions on $\Lambda^2$ as being applicable to the fibers of the vector bundle $\Lambda^2(M)$ of 2-forms on the spacetime manifold $M$, which is no longer assumed to necessarily be a vector space. The continuous distribution of oscillators that facilitates wave motion is then really a bundle of electromagnetic oscillators and an electromagnetic wave is a particular type of section of that bundle.

Some of the constructions that we just discussed are straightforward to generalize from vector spaces to vector bundles: When the each fiber of a vector bundle admits a complex structure, as a vector space, one says that the bundle itself admits an *almost-complex* structure. For $\Lambda^2(M)$, this becomes a vector bundle isomorphism *: $\Lambda^2(M) \to \Lambda^2(M)$ that takes each fiber to itself linearly and satisfies $*^2 = -I$. A decomposition of $\Lambda^2$ into real and imaginary subspaces becomes a decomposition of $\Lambda^2(M)$ into a Whitney sum $\Lambda^2_{\text{Re}}(M) \oplus \Lambda^2_{\text{Im}}(M)$ of vector bundles with three-dimensional fibers. A Hermitian structure on $\Lambda^2(M)$ is a Hermitian structure on each fiber, just a Lorentzian structure on $T(M)$ is a Minkowski space structure on each fiber.

The Hamiltonian function that is defined by the Hermitian structure becomes a function on $\Lambda^2(M)$ that still takes the local form of (4.4), except that one must understand that $E$ and **B** are covector and vector fields, respectively, on some open subset $U \subset M$ and no longer covectors and vectors. However, although the Kähler 2-form *K* makes each fiber of $\Lambda^2(M)$ into a symplectic vector space, it does not make the manifold $\Lambda^2(M)$ itself into a symplectic manifold, since its "vertical lift" to a 2-form on $\Lambda^2(M)$ is degenerate when one includes non-vertical tangent vectors. However, this does not prevent us from defining a vertical lift of the canonical Hamiltonian vector field that one obtains from the Hermitian form on each fiber to a vertical vector field on $\Lambda^2(M)$.

The complications associated with the extension from electromagnetic oscillators to electromagnetic wave motion arise when one wishes to introduce the notion that this continuous distribution of oscillators must also be "coupled," as well, even if only locally. That is because the differentials of the sections of the bundle $\Lambda^2(M)$ do not transform properly under changes of frame in the fibers. Although the introduction of the exterior derivative alleviates this to some degree, one should keep in mind that this invariance is $GL(4; \mathbb{R})$ frame invariance and is therefore limited to the 6-frames on $\Lambda^2$ that are due to 4-frames on $\mathbb{R}^4$ since this is consistent with the tensor product representation of $GL(4; \mathbb{R})$ in $GL(6; \mathbb{R})$, as it acts on $\Lambda^2$. Hence, if the more general 6-frames play an unavoidable role in the geometry that one must deal with when starting with to begin with, then possibly one must regard the associated principal bundle to the vector bundle $\Lambda^2(M)$ as the $GL(6; \mathbb{R})$-principal bundle $GL(6)(\Lambda^2)$ defined by *all* 6-frames on the fibers of $\Lambda^2(M)$ and not the $GL(4; \mathbb{R})$-principal bundle that is defined the 6-frames that are defined by the tensor product representation. In such an event, one would also have to introduce a $\mathfrak{gl}(6; \mathbb{R})$ connection on $GL(6)(\Lambda^2)$ and work with covariant differentials, as one does in the metric geometry of the tangent bundle.

This would enlarge the scope of the problem from projective geometry to projective differential geometry. Since a discussion of projective differential geometry is beyond the scope of the present work, and must be approached with a certain sense of caution from the physical standpoint, we will defer that discussion to later research.

## 7 Discussion

In addition to aforementioned the extension of the foregoing constructions and results from vector spaces to vector bundles there are some other directions of further development that are important from a physical standpoint:

One of the obvious physical limitations of the previous discussion was the fact that we confined ourselves from the outset to linear electromagnetic constitutive laws. Consequently, with the growing scope of nonlinear electrodynamics, and in particular, nonlinear optics, eventually one must pursue the extension from linear to nonlinear constitutive laws, which then leads to nonlinear electromagnetic oscillators, and nonlinear electromagnetic waves. Since a crucial notion in quantum electrodynamics is that of vacuum polarization, which represents a nonlinear contribution to the linear homogeneous isotropic constitutive law that is commonly associated with the classical electromagnetic vacuum, it is also conceivable that this extension might lead to a better understanding of what quantum electrodynamics has to say about the quantum electromagnetic vacuum.

Along similar lines, if one thinks of the transition from classical electromagnetic fields to quantum electromagnetic fields as relating to the replacement of the classical electromagnetic oscillators with quantum harmonic oscillators then this direction of research might prove illuminating, as well.

Finally, since the group *SU*(3) ordinarily first surfaces in physics in the context of the color gauge symmetry of the strong interaction it seems odd to see it emerge in such a classically familiar context as that of the anisotropic three-dimensional harmonic oscillator. One might even ponder the possible relationship between these two pictures. For instance, one must recall that strong interaction was first introduced as a way to stabilize the atomic nuclei with more than one proton in the face of the enormous Coulomb repulsion that was predicted by classical electrostatics. Interestingly, the same problem confronted early theorists who attempted to model the classical electron.